\documentclass[conference]{IEEEtran}
\IEEEoverridecommandlockouts
\usepackage{cite}
\usepackage{amsmath,amssymb,amsfonts}
\usepackage{algorithm2e}
\usepackage{graphicx}
\usepackage{textcomp}
\usepackage{xcolor}
\def\BibTeX{{\rm B\kern-.05em{\sc i\kern-.025em b}\kern-.08em
    T\kern-.1667em\lower.7ex\hbox{E}\kern-.125emX}}
\usepackage[binary-units]{siunitx}
\DeclareSIUnit{\bps}{bps}
\usepackage[caption=false,font=footnotesize]{subfig}
\usepackage{tabularx}
\usepackage{multirow}
\usepackage{flushend}

\begin{document}
\title{Initial Access for Millimeter-Wave and Terahertz Communications with Hybrid Beamforming}

\author{
	\IEEEauthorblockN{Wei~Jiang\IEEEauthorrefmark{1}\IEEEauthorrefmark{2},  and Hans~D.~Schotten\IEEEauthorrefmark{2}\IEEEauthorrefmark{1}}
	\IEEEauthorblockA{\IEEEauthorrefmark{1}Intelligent Networking Research Group, German Research Centre for Artificial Intelligence (DFKI), Germany\\
		\IEEEauthorrefmark{2}Institute for Wireless Communication and Navigation, Technische Universit\"at (TU) Kaiserslautern, Germany
	}%
	\thanks{}
}
\maketitle

\begin{abstract}
In order to achieve terabits-per-second (Tbps) data rates in the sixth-generation (6G) mobile system, wireless communications are required to exploit the abundant spectrum in the millimeter-wave (mmWave) and terahertz (THz) bands. However, high-frequency transmission heavily relies on high beamforming gain to compensate for severe propagation loss. A beam-based system faces a barrier in the process of initial access, where a base station must broadcast synchronization signals and system information to all users within its coverage. Hence, this paper proposes a novel omnidirectional broadcasting scheme for mmWave and THz systems with hybrid beamforming. It provides an instantaneously equal gain over all directions by forming complementary beams over sub-arrays.  Numerical results verify that it can achieve omnidirectional coverage with a performance that remarkably outperforms the previous scheme.
\end{abstract}

\section{Introduction}
Although the fifth-generation (5G) technology \cite{Ref_jiang2017experimental} is still on its way to being deployed across the world, academia and industry have already shifted their focus towards the next-generation technology known as the sixth generation (6G) \cite{Ref_jiang2021kickoff}.  The ITU-T focus group \textit{Technologies for Network 2030} envisioned that 6G will support disruptive applications, such as holographic communications, extended reality, artificial intelligence \cite{Ref_jiang2020deep, Ref_jiang2019recurrent, Ref_jiang2018multi},  Tactile Internet, multi-sense experience, and digital twin, which impose extreme capacities and performance requirements, including a peak data rate of 1 terabits-per-second (Tbps), a missive connection density of $10^7$ devices per \si{\kilo\meter^2}, and an area traffic capacity of \SI{1}{\giga\bps\per\meter^2}  \cite{Ref_jiang2021road}.

Millimeter-wave (mmWave) communications have been employed in 5G, but the identified spectrum is still very limited relative to the demand of 6G. To be specific, the ITU-R has assigned a total of \SI{13.5}{\giga\hertz} bandwidth for mmWave communications at the World Radiocommunication Conference 2019 (WRC-19). As a follow-up, 3GPP specified the Second Frequency Range (FR2) of 5G New Radio (NR), covering \SI{24.25}{\giga\hertz} to \SI{52.6}{\giga\hertz}. Using a bandwidth on the order of magnitude around \SI{10}{\giga\hertz}, a data rate of \SI{1}{\tera\bps} can only be achieved under a spectral efficiency approaching \SI{100}{\bps/\hertz}, which requires symbol fidelity that is not feasible using currently known digital modulation techniques or transceiver components. Consequently, 6G has to exploit the massively abundant spectrum at mmWave frequencies, which are defined as \SIrange{30}{300}{\giga\hertz}, and terahertz (THz) usually covering \SIrange{0.1}{3}{\tera\hertz}. At the WRC-19, the ITU-R has already identified the spectrum between \SI{275}{\giga\hertz} and \SI{450}{\giga\hertz} for the use of land mobile and fixed services, paving the way of deploying THz commutations in 6G \cite{Ref_kuerner2020impact}.

Despite its high potential, mmWave and THz communications suffer from severe propagation losses raised from high free-space path loss, atmospheric gaseous absorption, and weather attenuation \cite{Ref_siles2015atmospheric}, leading to a very short transmission range. Hence, antenna arrays are required at the base station and/or the mobile terminal to achieve sufficient beamforming gains to compensate for such losses. Nevertheless, a beam-based system encounters the problem of \textit{initial access} \cite{Ref_giordani2016initial, Ref_barati2016initial, Ref_li2017design}. In any cellular system, when a terminal powers on or performs the transition from the IDEL to CONNECTED mode, it needs to search for a suitable cell to access. Meanwhile, a terminal needs to detect the neighboring cells of its serving cell to prepare for handover. To this end, base stations must periodically broadcast synchronization signals and system information in the downlink with \textit{omnidirectional coverage}.

The traditional approach of transmitting synchronization and broadcast signals uses a single antenna that has an omnidirectional radiation pattern. Consequently, signal broadcasting was never a concern in earlier mobile systems that employed single-antenna base stations. However, mmWave and THz systems rely on pencil beams to transmit both control and user data, whereas beam-based broadcasting covering mobile users at any direction is challenging \cite{Ref_barati2016initial}.  The initial access of a 5G NR system is achieved by a brute-force method called \textit{beam sweeping} to sequentially scan the \SI{360}{\degree} angular space with multiple narrow beams \cite{Ref_dahlman20215gNR}. It applies a predefined codebook consisting of a set of weighting vectors, each of which forms a narrow beam to cover a particular direction, and all beams together seamlessly cover the whole angular space. 5G NR first defined the term Synchronization Signal Block (SSB), consisting of the Primary Synchronization Signal (PSS), Secondary Synchronization Signal (SSS), and Physical Broadcast Channel (PBCH). The set of SSBs within a beam-sweeping procedure to scan \SI{360}{\degree} is referred to as an \textit{SS burst set}. An NR system operating in the FR2 band can support up to $L=64$ SSBs within a burst set using \num{64} beams for sweeping. An SSB is transmitted over each beam to guarantee that all directions can receive synchronization signals and master system information.  This method provides good coverage but suffers from high overhead (the SSB is repeatedly transmitted $L$ times) and long discovery delay \cite{Ref_desai2014initial}.

At present, the most appropriate technique for initial access in mmWave and THz communications is called \textit{random beamforming} (RBF) proposed by Yang and Jiang through \cite{Ref_yang2013random, Ref_yang2012methodUS8170132, Ref_jiang2012methodUS13685426, Ref_yang2013methodUS8537785, Ref_jiang2012enhanced, Ref_yang2012methodUS13654743}. But it still has some drawbacks. Hence, this paper proposes a novel initial-access technique called \textit{complementary beamforming} (CBF), which goes beyond the RBF in terms of the following aspects:
\begin{itemize}
    \item The RBF is based on \textit{digital beamforming}, which is too expensive and power-consuming for large-scale arrays. Our proposal focuses on hybrid beamforming that is more suitable for mmWave and THz communications. It makes full use of sub-arrays in the hybrid digital-analog architecture to form complementary patterns.
    \item The RBF achieves omnidirectional coverage by \textit{averaging} many beams over the time or frequency domain. There is a performance loss compared to the benchmark (the single-antenna broadcasting) due to the power fluctuation in the angular domain. In contrast, the CBF can provide \textit{instantaneously} isotropic gain using complementary beams and thus improve performance.
    \item The RBF needs to form many random beam patterns in a short time/frequency span to achieve averaged equal gains over all directions. The isotropic gain of the CBF is achieved by a pair of complementary beams, which can be fixed. Thus, it simplifies the implementation.
\end{itemize}

The remainder of this paper is organized as follows: Section II introduces the system model, while Section III proposes complementary beamforming. Finally, Section IV gives numerical results, and conclusions are made in Section V.

\section{System Model}
Implementing digital beamforming in a mmWave or THz transceiver equipped with a large-scale array needs a large number of RF components, leading to high hardware cost and power consumption. This constraint has driven the application of analog beamforming, using only a single RF chain. Analog beamforming is implemented as the \textit{de-facto} approach for indoor mmWave systems. However, it only supports single-stream transmission and suffers from the hardware impairment of analog phase shifters, e.g., random phase noise \cite{Ref_jiang2021impactcellfree}. As a consequence, hybrid beamforming \cite{Ref_zhang2019hybrid} has been proposed as an efficient approach to support multi-stream transmission with only a few RF chains and a phase-shifter network. Compared with analog beamforming, hybrid beamforming supports spatial multiplexing, diversity, and spatial-division multiple access. It achieves spectral efficiency comparable to digital beamforming with much lower hardware complexity and cost. Therefore, it is promising for mmWave and THz transceivers in 6G systems \cite{Ref_ahmed2018survey}.

\begin{figure}[!bpht]
\centering
\includegraphics[width=0.42\textwidth]{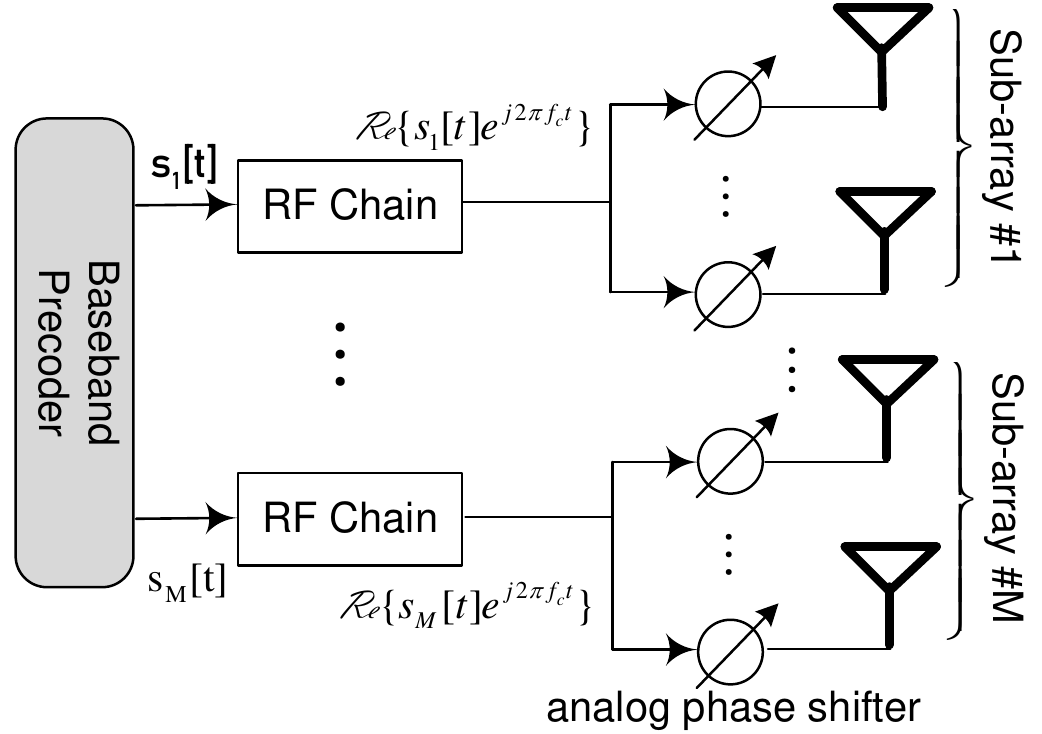}
\caption{Block diagrams of partially-connected hybrid beamforming. }
\label{Diagram_hybridBF}
\end{figure}

Hybrid beamforming can be implemented with different forms of circuit networks, resulting in two basic structures:
\begin{itemize}
    \item \textit{Fully-Connected Hybrid Beamforming}\\ The transmit data is first precoded in the baseband into $M$ data streams, and each stream is processed by an independent RF chain. Each RF chain is connected to all $N$ antennas via an analog network, where $N\gg M$. Hence, a total of $MN$ analog phase shifters are required.
    \item \textit{Partially-Connected Hybrid Beamforming} \\ As shown in \figurename \ref{Diagram_hybridBF}, each RF chain is only connected to a subset of all antenna elements called a sub-array. This structure is preferred from the perspective of practice since it substantially lowers hardware complexity (as well as power consumption) by dramatically reducing the number of analog phase shifters from $MN$ to $N$.
\end{itemize}

Without loss of generality, we use partially-connected hybrid beamforming hereinafter to analyze the proposed scheme but its applicability to fully-connected hybrid beamforming will be discussed in Sub-Sec. \ref{SecExtension}. Mathematically, the output of the baseband precoder is denoted by $s_m[t]$, $1\leq m \leq M$, as shown in \figurename \ref{Diagram_hybridBF}. After digital-to-analog conversion and up-conversion, the $m^{th}$ RF chain feeds
\begin{equation}
    \Re\left [s_m[t]e^{j2\pi f_0 t}\right], \:\:1\leq m \leq M,
\end{equation}
into the analog network, where $\Re [\cdot]$ denotes the real part of a complex number, and $f_0$ stands for the carrier frequency.  In the partially-connected beamforming, an array is divided into several sub-arrays, and each antenna is allocated to only one RF chain. Each sub-array contains $N_s=N/M$ elements (suppose $N_s$ is an integer).  We write $\phi_{nm}$, for $1\leq n \leq N_s$ and $1\leq m \leq M$ to denote the added phase shift on the $n^{th}$ antenna of the $m^{th}$ sub-array, and thus the vector of transmit signals for the $m^{th}$ sub-array is given by
\begin{align} \label{eqn:mmWave:TxSingalPerSubarray}
    &\textbf{s}_m(t) = \\ \nonumber
    &\biggl[\Re\left [s_m[t]e^{j2\pi f_0t}e^{j\phi_{1m}}\right],\ldots,\Re\left [s_m[t]e^{j2\pi f_0t}e^{j\phi_{N_sm}}\right] \biggr]^T.
\end{align}

Radiating a plane wave into a homogeneous media in the direction indicated by the angle of departure $\theta$, the time difference between a typical element $n$ of the $m^{th}$ sub-array and the reference point is denoted by $\tau_{nm}(\theta)$.
Within a flat-fading wireless channel, the received passband signal is
\begin{align}  \label{eqn:mmWave:RXsignalPatiallyHybridBF} \nonumber
    &y(t)  =  \sum_{m=1}^M \sum_{n=1}^{N_s} \Re\left [h_m(t)s_m[t]e^{j2\pi f_0(t-\tau_{nm}(\theta))}e^{j\phi_{nm}}\right] + n(t)\\ \nonumber
         &=  \Re\left [ \left(\sum_{m=1}^M h_m(t)s_m[t] \sum_{n=1}^{N_s}  e^{-j2\pi f_0\tau_{nm}(\theta)}e^{j\phi_{nm}}\right) e^{j2\pi  f_0t} \right]\\
         &+ n(t),
\end{align}
where $h_m(t)$ represents the channel response between the $m^{th}$ sub-array and receiver, and $n(t)$ the noise. In wideband communications, a frequency-selective channel can be transferred to a set of frequency-flat channels, where the scheme can be straightforwardly applied over each subcarrier \cite{Ref_jiang2016ofdm}.

Denoting the steering vector of sub-array $m$ as
\begin{equation}
    \textbf{a}_m(\theta) = \left[e^{-j2\pi f_0\tau_{1m}(\theta)},\ldots,e^{-j2\pi f_0\tau_{N_sm}(\theta)}   \right]^T,
\end{equation}
and its weighting vector (due to analog phase shift) as
\begin{equation}
    \textbf{w}_m = \left[e^{j\phi_{1m}}, \ldots, e^{j\phi_{N_sm}}   \right]^T,
\end{equation}
Eq. (\ref{eqn:mmWave:RXsignalPatiallyHybridBF}) can be rewritten as
\begin{align} \nonumber \label{eqn:mmWave:BeamPatternPahyBF}
    y(t) &=   \Re\left [\left( \sum_{m=1}^M h_m(t) s_m[t] \textbf{w}_m^T \textbf{a}_m(\theta)  \right)e^{j2\pi  f_0t}\right] +n(t),\\
    &=\Re\left [\left( \sum_{m=1}^M h_m(t) s_m[t] g_m(\theta,t) \right)e^{j2\pi  f_0t}\right] +n(t),
\end{align}
with the beam pattern generated by the $m^{th}$ sub-array:
\begin{equation} \label{eqn:ICC:beampattern}
   g_m(\theta,t)=\textbf{w}_m^T \textbf{a}_m(\theta)=\sum_{n=1}^{N_s}  e^{-j2\pi f_0\tau_{nm}(\theta)}e^{j\phi_{nm}}.
\end{equation} After down-conversion and sampling at the receiver, the baseband equivalent received signal can be simplified into (neglecting the time index)
\begin{equation} \label{eqn:ICC:BBRxSignal}
    y =  \sum_{m=1}^M  h_m g_m(\theta) s_m  +n,
\end{equation}
where $y$ is the received symbol, $s_m$ is the precoded symbol for the $m^{th}$ RF chain, and $h_m$ denotes the baseband channel coefficient between the $m^{th}$ sub-array and receiver.

\section{Complementary Beamforming}
\subsection{Design Criteria}
In the previous RBF scheme \cite{Ref_yang2011random}, the time-frequency resources are divided into many small time-frequency blocks (TFB) indexed by $t$, and a random pattern is applied on each TFB. For a sufficient number of random patterns, the average power is nearly equal for each direction. Then, the RBF achieves omnidirectional coverage with \textit{averaged} equal gain at any direction.
The criteria in designing the random pattern sequence can be summarized as follows
\begin{itemize}
    \item Keep equal average power in each direction for omnidirectional coverage
    \item Set equal power in each antenna to maximize power amplifier efficiency
    \item Use random beams with the minimum variance to achieve the maximal capacity
\end{itemize}
To achieve the objective of isotropic radiation, \textit{the pattern variance} in the angular dimension is defined as a metric to measure the pattern's degree of deviation from a circle:
\begin{equation} \label{eqn:beamVariation}
    \sigma_g^2 = \frac{1}{2\pi}\int_0^{2\pi} \biggl[ \bigl |g(\theta,t)\bigr|^2 - \mathbb{E}\left( \bigl|g(\theta,t)\bigr|^2\right) \biggr]^2 d\theta,
\end{equation}
where $\mathbb{E}$ means mathematical expectation.
To satisfy the second and third criteria, the basis weighting vector has the minimum variance and unit module for each entry, i.e.,
\begin{itemize}
   \item $\textbf{w}_0=\arg\min (\sigma_g^2)$,
    \item $|w_1|=|w_2|=\dots=|w_N|=1$.
\end{itemize}

\subsection{Complementary Patterns in Hybrid Beamforming}
Despite achieving omnidirectional coverage with averaged equal power, the power distribution of an individual random beam in the RBF still fluctuates in the angular domain, as shown in Fig. 2 of \cite{Ref_yang2013random}. Compared with the single-antenna broadcasting with even energy distribution, this fluctuation leads to performance loss. To be specific, a beam null such as that located in \SI{0}{\degree} or \SI{180}{\degree} brings a low signal-to-noise ratio and then error bits, analog to a deep fade in a wireless channel. However, it is impossible to form a \textit{circular} beam over some types of arrays, such as a uniform linear array (ULA). To avoid this constraint, this paper proposes to use a pair of beams (or more), taking advantage of sub-arrays in hybrid beamforming, rather than a single beam. These two beams are complementary to each other to form omnidirectional coverage with \textit{instantaneous} equal gain at any direction. As a result, the optimal performance identical to that of a single antenna with a high PA can be obtained.

Without loss of generality, we use hybrid beamforming with two RF chains, i.e., $M=2$, as an example, while its generalization will be discussed in Sub-Sec. \ref{SecExtension}.
Instead of a unique weighting vector in the RBF, the CBF determines a pair of weighting vectors, satisfying the following criteria:
\begin{itemize}
     \item Minimize the variance of individual pattern, i.e.,
        \begin{equation} \nonumber \hat{\textbf{w}}_k=\arg\min(\sigma^2_{g_k}),\: k=1,2, \end{equation}
        where $g_k$ denotes the individual pattern of $\textbf{w}_1$ and $\textbf{w}_2$ over sub-array \num{1} and \num{2}, respectively.
     \item Minimize the variance of the composite pattern, i.e.,
        \begin{equation} \nonumber \bigl\{\hat{\textbf{w}}_1,\hat{\textbf{w}}_2\bigr\}=\arg\min\left(\sigma^2_{g}\right). \end{equation}
        We define the amplitude of the composite pattern as  \begin{equation}
        |g(\theta,t)| =\sqrt{\frac{|g_1( \theta,t)|^2+|g_2(\theta,t)|^2}{2}}.
    \end{equation}
    \item Set equal transmit power in each antenna to maximize the PA efficiency, i.e., \begin{equation}  |w_1|=|w_2|=\dots=|w_{N_s}|=1.\end{equation}
\end{itemize}

\begin{figure}[!bpht]
\centering
\includegraphics[width=0.5\textwidth]{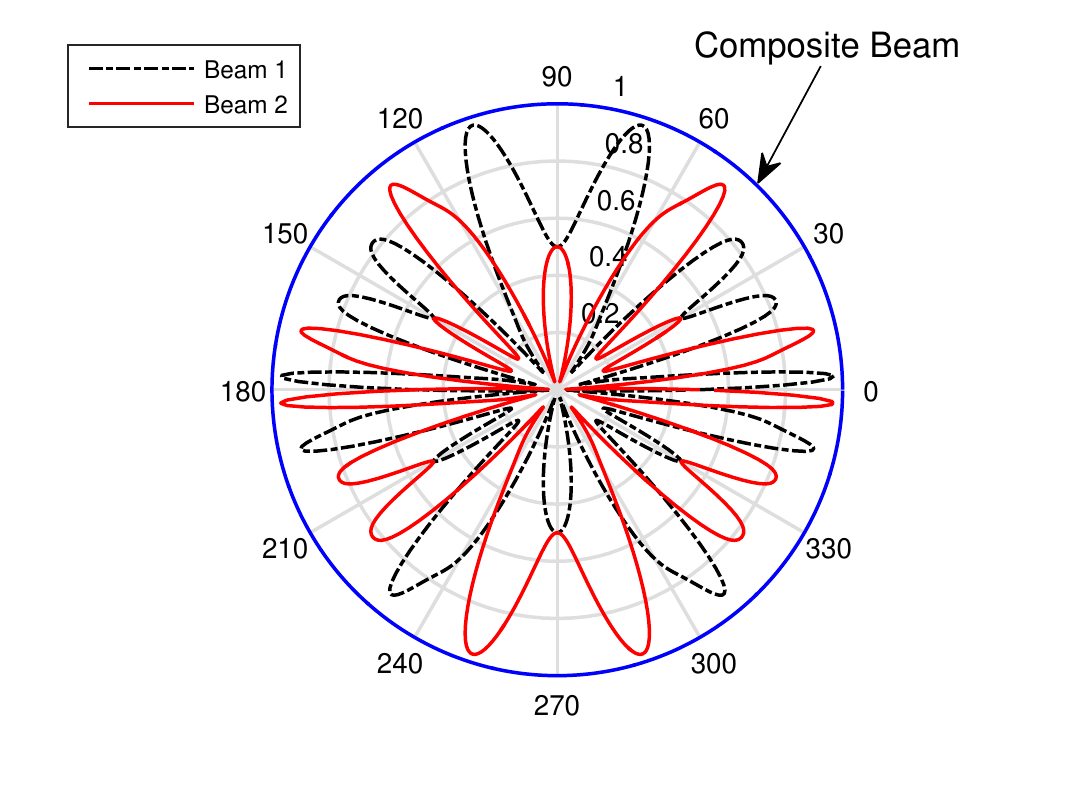}
\caption{Schematic diagram of a pair of complementary beams over a 16-element ULA, generating a composite pattern with instantaneously equal gain over all directions, satisfying ($\sigma_g^2=0$).  }
\label{Diagram_BeamPair}
\end{figure}

Given the steering vectors of sub-arrays, the desired weighting vectors can be determined by conducting a computer search, as depicted in \algorithmcfname~\ref{alg:001}.
This process is not computationally complex, even with large antenna numbers. Moreover, engineers can figure out these weighting vectors during system design and configure equipment beforehand, imposing no burden on the system operation.  \figurename \ref{Diagram_BeamPair} illustrates a pair of complementary beams over an 16-element ULA as an example. Although either beam still fluctuates in the angular domain, their composite pattern is strictly isotropic with zero deviation $\sigma_g^2=0$, namely
\begin{equation}
     |g|^2=\frac{|g_1|^2+|g_2|^2}{2}=1.
\end{equation}

\SetKwComment{Comment}{/* }{ */}
\RestyleAlgo{ruled}
\begin{algorithm}
\caption{Determine Complementary Beams} \label{alg:001}
\SetKwInOut{Input}{input}\SetKwInOut{Output}{output} \SetKwInput{kwInit}{Initialization}
\Input{$\textbf{a}_m(\theta)$, $m=1,2$}
\Input{Accuracy factor $K>1$}
\KwResult{$\tilde{\mathbf{w}}_1$,$\tilde{\mathbf{w}}_2$}
Coefficient granularity $\vartriangle\phi \gets \frac{2\pi}{K}$\;
A coefficient set $ \mathbf{C} \gets \{1,e^{\vartriangle \phi},e^{2\vartriangle \phi},\ldots,e^{(K-1)\vartriangle \phi} \}$\;
Space of weighting vectors $\mathbf{W} \gets \mathbf{C}^{N_s}$\;
\kwInit { $\tilde{\sigma}^2_g \gets 1$, $\tilde{\mathbf{w}}_1\gets \mathbf{0}$, $\tilde{\mathbf{w}}_2\gets \mathbf{0}$}
\ForEach{$\mathbf{w}_1, \mathbf{w}_2 \in \mathbf{W}$}{
  $g_1(\theta)\gets\mathbf{w}^T_1\textbf{a}(\theta)$,
  $g_2(\theta)\gets\mathbf{w}^T_2\textbf{a}(\theta)$\;
  $|g(\theta)|\gets \sqrt{\frac{|g_1(\theta)|^2+|g_2(\theta)|^2}{2}}$\;
  Compute $\sigma^2_g$ in terms of Eq. (\ref{eqn:beamVariation})\;
  \If{$\sigma^2_g<\tilde{\sigma}^2_g$}{
    $\tilde{\sigma}^2_g \gets \sigma^2_g$,
    $\tilde{\mathbf{w}}_1 \gets \mathbf{w}_1$,
    $\tilde{\mathbf{w}}_2 \gets \mathbf{w}_2$\;
  }
}
\end{algorithm}

\subsection{Baseband Precoding and Decoding}

\begin{figure}[!bpht]
\centering
\includegraphics[width=0.42\textwidth]{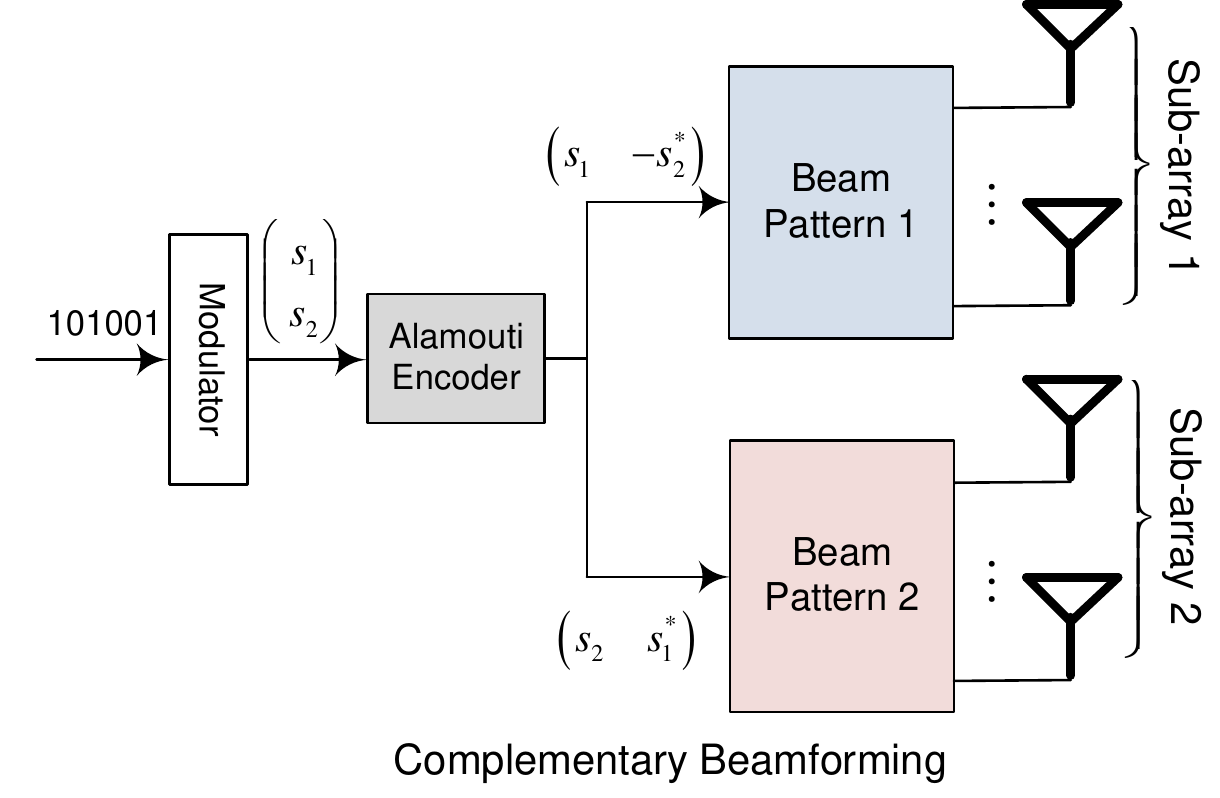}
\caption{Schematic diagram of complementary beamforming over partially-connected hybrid transmitter, where two branches of orthogonal precoded signals are independently beam formed over two sub-arrays to generate two complementary patterns. }
\label{Diagram_complementaryBeamformer}
\end{figure}

In digital beamforming, a data symbol is weighted correctly and transmitted at all elements to achieve constructively superposition in desired directions. If two sub-arrays of hybrid beamforming transmit the same symbol, i.e., $s_1=s_2$, and their channels are fully correlated $h_1=h_2=h$, Eq. (\ref{eqn:ICC:BBRxSignal}) becomes
\begin{equation}
    y =   \bigl(g_1(\theta)+ g_2(\theta)\bigr)hs_m  +n.
\end{equation}
In terms of Eq. (\ref{eqn:ICC:beampattern}), we can obtain
\begin{align} \nonumber
      g_1(\theta)+ g_2(\theta)&=\sum_{m=1}^{2}\sum_{n=1}^{N_s}  e^{-j2\pi f_0\tau_{nm}(\theta)}e^{j\phi_{nm}}\\ \nonumber
      &=\sum_{n'=1}^{2N_s}  e^{-j2\pi f_0\tau_{n'm}(\theta)}e^{j\phi_{n'm}}\\
      &=g(\theta,t),
\end{align}
implying that hybrid beamforming falls back to analog beamforming. It imposes the requirements that different streams of hybrid beamforming should be independent to form individual patterns. Meanwhile, the broadcasting of synchronization signals and system information emphasize on reliability, rather than throughput. To this end, the precoding scheme with spatial diversity is preferred. The Alamouti scheme \cite{Ref_alamouti1998simple}, is the unique complex-symbol space-time block code (STBC) with full transmit diversity at full symbol rate, which has been applied in practical systems and is quite flexible to be integrated with other schemes, e.g., opportunistic space-time coding \cite{Ref_jiang2014opportunistic}.

A pair of symbols $\mathbf{s}=[s_1,s_2]^T$ are precoded as follows:
\begin{equation}  \begin{pmatrix}
 s_1 \\
 s_2
\end{pmatrix} \xrightarrow{\text{Precoding}}
    \begin{pmatrix}
 s_1& -s_2^* \\
 s_2& s_1^*
\end{pmatrix}.
\end{equation}
The row of the precoded matrix corresponds to the spatial domain, i.e., different sub-arrays, while the column stands for the temporal domain. In other words, $s_1$ and $-s_2^*$ are transmitted by sub-array 1 over two consecutive symbol periods, and meanwhile $s_2$ and $s_1^*$ are transmitted by sub-array 2, as shown in \figurename \ref{Diagram_complementaryBeamformer}.
In the complementary beamforming, two streams are independently beam formed over sub-arrays \num{1} and \num{2}, respectively. The electromagnetic interference phenomenon occurs only among elements transmitting correlated signals. Thereby, the pattern of different sub-arrays can be regarded as independent.
According to Eq. (\ref{eqn:ICC:BBRxSignal}), the received symbols over two consecutive symbol periods can be expressed by
\begin{equation} \label{eqn:ICC:Rxsignals}
    \left\{ \begin{aligned} y_1&= g_1h_1s_1 + g_2h_2s_2 +n_1   ,\\  y_2&= -g_1h_1s_2^* + g_2h_2s_1^* +n_2.\end{aligned}         \right.
\end{equation}
Building a vector of received symbols as $\mathbf{y}=[y_1,y_2^*]^T$, a noise vector $\mathbf{n}=[n_1,n_2^*]^T$, and a composite channel matrix
\begin{equation}  \mathbf{H}=
    \begin{pmatrix}
 g_1h_1& g_2h_2 \\
 g_2^*h_2^*& -g_1^*h_1^*
\end{pmatrix},
\end{equation}
which is assumed to be perfectly known from channel estimation, Eq. (\ref{eqn:ICC:Rxsignals}) is rewritten into matrix form as
\begin{equation}
    \mathbf{y} = \mathbf{H}\mathbf{s}+\mathbf{n}.
\end{equation}
With typical decoding schemes for the Alamouti coding, the transmitted symbols can be detected, e.g., using the minimum mean-square error (MMSE) decoding
\begin{equation}
    \hat{\mathbf{s}} = \left(\mathbf{H}^H\mathbf{H}+\sigma^2\mathbf{I}\right)^{-1}\mathbf{H}^H\mathbf{y},
\end{equation}
where $\sigma^2$ is the noise variance, and $\mathbf{I}$ denotes the unit matrix.

\subsection{Extension} \label{SecExtension}
In addition to partially-connected hybrid beamforming with two sub-arrays, which is applied as an example for simplicity, the proposed scheme is also applicable to fully-connected hybrid beamforming and any number of $M$ RF chains. The output of the procoder is fed into the RF chains, and each RF chain beam formed its corresponding transmitted signal over all elements. If we treat the whole array as same as a sub-array, it results in multiple complementary beams in fully-connected hybrid beamforming.
There are at least two RF chains in hybrid architecture. If the number of RF chains is even, we can group them into pairs, and apply the CBF directly  pair-by-pair. Two  sub-arrays corresponding to each pair of RF chains generates a pair of complementary beams. If the number of RF chains is odd, we can group the last three RF chains into one group, and then find three complementary beams to minimize the deviation of their composite pattern.


\begin{figure*}[!tbph]
\centerline{\hspace{0mm}
\subfloat[]{\includegraphics[width=0.4\textwidth]{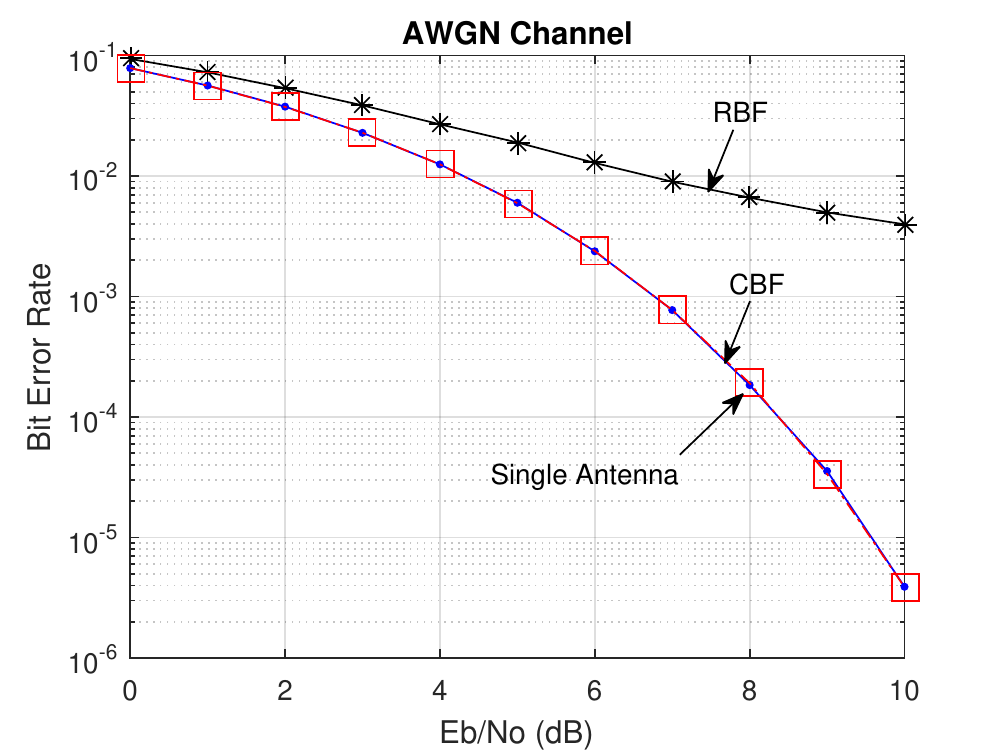}\label{fig:result1}} \\
\subfloat[]{\includegraphics[width=0.4\textwidth]{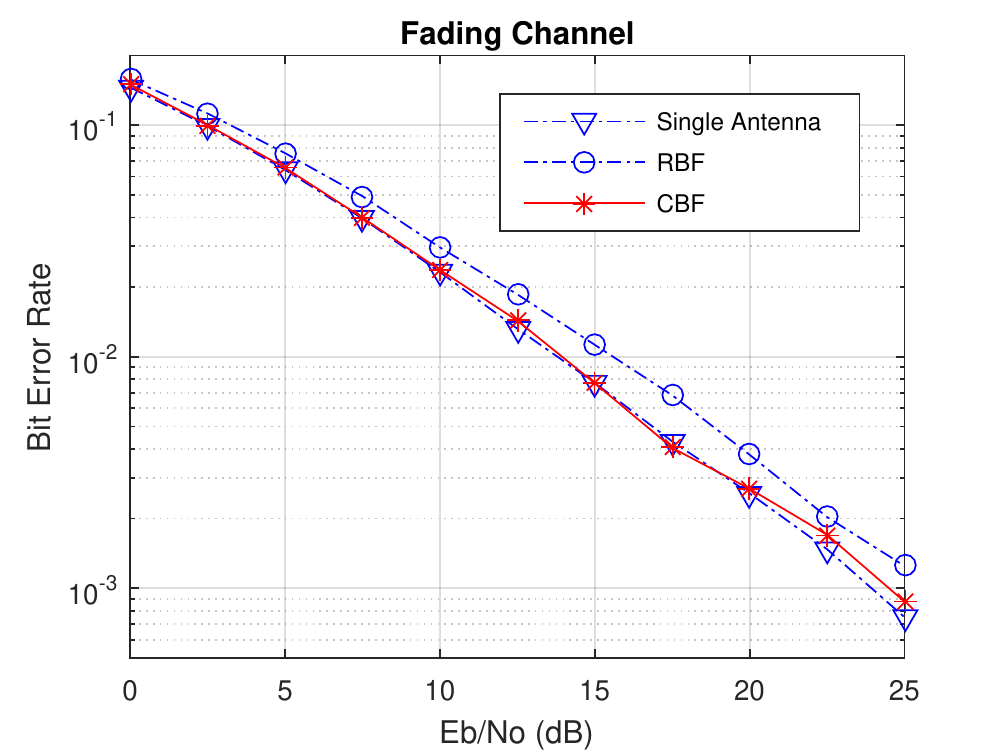}\label{fig:result2}} }
\hspace{0mm}
 \caption{Comparison of uncoded BER performance among the CBF, RBF, and single-antenna broadcasting in (a) Additive White  Gaussian Noise (AWGN) channels and (b) Frequency-flat Rayleigh fading channels.  }
\label{Diagram_Results}
\end{figure*}

\section{Simulations}
The performance of the proposed CBF in terms of the bit error rate (BER) is verified through numerical results acquired by computer simulations. Since link reliability is more important than the data rate in broadcasting synchronization signals and system information, low-order modulation, i.e., quadrature phase-shift-keying (QPSK), is employed. To get an insight into the performance without any other influential factors, the uncoded BER comparison in Additive White Gaussian Noise (AWGN) channels is conducted. Moreover, the results over Rayleigh fading channels are observed, where channel coefficients vary in block-wise.  For a direct comparison with the RBF scheme, an 8-element ULA is applied. We also assume that the channels of two sub-arrays are the same $h_1=h_2$. Then, the performance of the Alamouti precoding over two antennas is identical to that of the single antenna (as the benchmark) under the same power budget.

The proposed scheme is compared with the RBF, using the signal-antenna broadcasting as the benchmark.  Note that the single antenna needs a high power amplifier, which brings high hardware costs and power consumption. Moreover, mmWave and THz communications must rely on beamforming over a large-scale array to extend the communication range for transmitting both user data and control signalling. Although a single-antenna broadcasting scheme is optimal for synchronization and broadcast signals, it is not an option for user data transmission in mmWave and THz communications. Hence, it can only serve as the benchmark for omni-directional coverage, rather than a real candidate. In order to verify the omnidirectional coverage property of the CBF, we select different angles for a mobile station to observe its BER performance in the downlink. As we expect for the broadcast channels, the proposed scheme gets identical performance in any angle, as the RBF and the single-antenna broadcasting. However, the CBF outperforms the RBF since it provides instantaneous equal gains in all directions and avoids the fluctuation of energy distribution in the angular domain.

The total power budget for the three schemes is the same for a fair comparison. The uncoded BER performance in an AWGN and frequency-flat Rayleigh fading channel is illustrated in \figurename \ref{Diagram_Results}. The CBF can achieve the optimal performance indicated by the single antenna in both channels since it provides instantaneous equal gain without any beam null. The bit errors of the RBF in AWGN is raised from the beam nulls that cause a very low signal-to-noise ratio. In fading channels, all three schemes suffer from channel fading, where most of the bit errors are caused by the deep fade of wireless channels, but the CBF is also superior to the RBF.

\section{Conclusion}
This paper proposed a novel beamforming scheme for the initial access of future 6G systems, which adopt millimeter-wave and terahertz communications with hybrid analog-digital architecture. The proposed scheme provides omnidirectional coverage for the broadcasting of synchronization signals and master system information, such as Synchronization Signal Block specified in 5G New Radio, to all mobile uses at any angle of a cell. Unlike the previous RBF technique that achieves omnidirectional coverage by averaging many random patterns, it can generate instantaneously isotropic radiation to avoid the energy fluctuation in the angular domain by forming complementary beams. Consequently, it outperforms random beamforming since the performance loss is caused by the low signal-to-noise ratio due to beam nulls.

\bibliographystyle{IEEEtran}
\bibliography{IEEEabrv,Ref_ICC2022}

\begin{thebibliography}{10}
\providecommand{\url}[1]{#1}
\csname url@samestyle\endcsname
\providecommand{\newblock}{\relax}
\providecommand{\bibinfo}[2]{#2}
\providecommand{\BIBentrySTDinterwordspacing}{\spaceskip=0pt\relax}
\providecommand{\BIBentryALTinterwordstretchfactor}{4}
\providecommand{\BIBentryALTinterwordspacing}{\spaceskip=\fontdimen2\font plus
\BIBentryALTinterwordstretchfactor\fontdimen3\font minus
  \fontdimen4\font\relax}
\providecommand{\BIBforeignlanguage}[2]{{%
\expandafter\ifx\csname l@#1\endcsname\relax
\typeout{** WARNING: IEEEtran.bst: No hyphenation pattern has been}%
\typeout{** loaded for the language `#1'. Using the pattern for}%
\typeout{** the default language instead.}%
\else
\language=\csname l@#1\endcsname
\fi
#2}}
\providecommand{\BIBdecl}{\relax}
\BIBdecl

\bibitem{Ref_jiang2017experimental}
W.~Jiang \emph{et~al.}, ``Experimental results for {Artificial
  Intelligence}-based self-organized {5G} networks,'' in \emph{Proc. {IEEE}
  Int. Symp. on Pers., Indoor and Mobile Radio Commun. (PIMRC)}, Montreal,
  Canada, Oct. 2017.

\bibitem{Ref_jiang2021kickoff}
W.~Jiang and H.~D. Schotten, ``The kick-off of {6G} research worldwide: An
  overview,'' in \emph{Proc. 2021 Seventh IEEE Int. Conf. on Comput. and
  Commun. (ICCC)}, Chengdu, China, Dec. 2021.

\bibitem{Ref_jiang2020deep}
------, ``Deep learning for fading channel prediction,'' \emph{IEEE Open J. the
  Commun. Society}, vol.~1, pp. 320--332, Mar. 2020.

\bibitem{Ref_jiang2019recurrent}
W.~Jiang and H.~Schotten, ``Recurrent neural network-based frequency-domain
  channel prediction for wideband communications,'' in \emph{Proc. {IEEE} Veh.
  Tech. Conf. (VTC)}, Kuala Lumpur, Malaysia, Apr. 2019.

\bibitem{Ref_jiang2018multi}
W.~Jiang and H.~D. Schotten, ``Multi-antenna fading channel prediction
  empowered by artificial intelligence,'' in \emph{Proc. {IEEE} Veh. Tech.
  Conf. (VTC)}, Chicago, USA, Aug. 2018.

\bibitem{Ref_jiang2021road}
W.~Jiang \emph{et~al.}, ``The road towards {6G}: A comprehensive survey,''
  \emph{IEEE Open J. Commun. Society}, vol.~2, pp. 334--366, Feb. 2021.

\bibitem{Ref_kuerner2020impact}
T.~Kuerner and A.~Hirata, ``On the impact of the results of {WRC 2019 on THz}
  communications,'' in \emph{Proc. of 2020 Third Intl. Workshop on Mobile
  Terahertz Syst. (IWMTS)}, Essen, Germany, Jul. 2020, pp. 1--3.

\bibitem{Ref_siles2015atmospheric}
G.~A. Siles, J.~M. Riera, and P.~G. del Pino, ``Atmospheric attenuation in
  wireless communication systems at millimeter and {THz} frequencies,''
  \emph{{IEEE} Antennas Propag. Mag.}, vol.~57, no.~1, pp. 48 -- 61, Feb. 2015.

\bibitem{Ref_giordani2016initial}
M.~Giordani, M.~Mezzavilla, and M.~Zorzi, ``Initial access in {5G mmWave}
  cellular networks,'' \emph{{IEEE} Commun. Mag.}, vol.~54, pp. 40 -- 47, Nov.
  2016.

\bibitem{Ref_barati2016initial}
C.~N. Barati \emph{et~al.}, ``Initial access in millimeter wave cellular
  systems,'' \emph{{IEEE} Trans. Wireless Commun.}, vol.~15, no.~12, pp. 7926
  -- 7940, Sep. 2016.

\bibitem{Ref_li2017design}
Y.~Li \emph{et~al.}, ``Design and analysis of initial access in millimeter wave
  cellular networks,'' \emph{{IEEE} Trans. Wireless Commun.}, vol.~16, no.~10,
  pp. 6409 -- 6425, Oct. 2017.

\bibitem{Ref_dahlman20215gNR}
E.~Dahlman, S.~Parkvall, and J.~Sköld, \emph{{5G NR - The Next Generation
  Wireless Access Technology}}.\hskip 1em plus 0.5em minus 0.4em\relax London,
  the United Kindom: Academic Press, Elsevier, 2021.

\bibitem{Ref_desai2014initial}
V.~Desai \emph{et~al.}, ``Initial beamforming for {mmWave} communications,'' in
  \emph{Proc. of 2014 Asilomar Conf. on Signals, Syst. and Comput.}, Pacific
  Grove, USA, Nov. 2014, pp. 1926--1930.

\bibitem{Ref_yang2013random}
X.~Yang, W.~Jiang, and B.~Vucetic, ``A random beamforming technique for
  omnidirectional coverage in multiple-antenna systems,'' \emph{{IEEE} Trans.
  Veh. Technol.}, vol.~62, no.~3, pp. 1420 -- 1425, Mar. 2013.

\bibitem{Ref_yang2012methodUS8170132}
X.~Yang and W.~Jiang, ``Method and apparatus for transmitting signals in a
  multiple antennas system,'' U.S. Patent 8\,170\,132, May 1, 2012.

\bibitem{Ref_jiang2012methodUS13685426}
W.~Jiang and X.~Yang, ``Method and apparatus for transmitting broadcast
  signal,'' U.S. Patent Application 13/685\,426, Nov. 26, 2012.

\bibitem{Ref_yang2013methodUS8537785}
X.~Yang and W.~Jiang, ``Method and apparatus for cell/sector coverage of a
  public channel through multiple antennas,'' U.S. Patent 8\,537\,785, Sep. 17,
  2013.

\bibitem{Ref_jiang2012enhanced}
W.~Jiang and X.~Yang, ``An enhanced random beamforming scheme for signal
  broadcasting in multi-antenna systems,'' in \emph{Proc. 2012 IEEE 23rd Int.
  Symp. on Pers., Indoor and Mobile Radio Commun. (PIMRC)}, Sydney, NSW,
  Australia, Sep. 2012, pp. 2055 -- 2060.

\bibitem{Ref_yang2012methodUS13654743}
X.~Yang and W.~Jiang, ``Method, apparatus, and system for controlling
  multi-antenna signal transmission,'' U.S. Patent Application 13/654\,743,
  Oct. 18, 2012.

\bibitem{Ref_jiang2021impactcellfree}
W.~Jiang and H.~Schotten, ``Impact of channel aging on zero-forcing precoding
  in cell-free massive {MIMO} systems,'' \emph{{IEEE} Commun. Lett.}, vol.~25,
  no.~9, pp. 3114 -- 3118, Sep. 2021.

\bibitem{Ref_zhang2019hybrid}
J.~Zhang \emph{et~al.}, ``Hybrid beamforming for {5G} and beyond
  millimeter-wave systems: A holistic view,'' \emph{IEEE Open J. Commun.
  Society}, vol.~1, pp. 77 -- 91, 12 2019.

\bibitem{Ref_ahmed2018survey}
I.~Ahmed \emph{et~al.}, ``A survey on hybrid beamforming techniques in {5G}:
  Architecture and system model perspectives,'' \emph{IEEE Commun. Surv.
  Tutor.}, vol.~20, no.~4, pp. 3060 -- 3097, 2018, fourthquarter.

\bibitem{Ref_jiang2016ofdm}
W.~Jiang and T.~Kaiser, ``From {OFDM} to {FBMC}: Principles and
  {Comparisons},'' in \emph{Signal Processing for 5G: Algorithms and
  Implementations}, F.~L. Luo and C.~Zhang, Eds.\hskip 1em plus 0.5em minus
  0.4em\relax United Kindom: John Wiley\&Sons and IEEE Press, 2016, ch.~3.

\bibitem{Ref_yang2011random}
X.~Yang, W.~Jiang, and B.~Vucetic, ``A random beamforming technique for
  broadcast channels in multiple antenna systems,'' in \emph{Proc. of 2011 IEEE
  Veh. Techno. Conf. (VTC Fall)}, San Francisco, USA, Sep. 2011, pp. 1--6.

\bibitem{Ref_alamouti1998simple}
S.~Alamouti, ``A simple transmit diversity technique for wireless
  communications,'' \emph{{IEEE} J. Sel. Areas Commun.}, vol.~16, no.~8, pp.
  1451 -- 1458, Oct. 1998.

\bibitem{Ref_jiang2014opportunistic}
W.~Jiang, H.~Cao, and T.~Kaiser, ``Opportunistic space-time coding to exploit
  cooperative diversity in fast-fading channels,'' in \emph{Proc. of 2014 IEEE
  Int. Commun. Conf. (ICC)}, Sydney, NSW, Australia, Jun. 2014, pp. 4814--4819.

\end{thebibliography}

\end{document}